# Molecular Dynamics Study of Bamboo-like Carbon Nanotube Nucleation

Feng Ding[1*], Kim Bolton[1,2] and Arne Rosén[1]

1) Department of Physics, Göteborg University, SE-412 96, Göteborg, Sweden

2) School of Engineering, University College of Borås, SE-50190, Borås, Sweden

**Abstract**

MD simulations based on an empirical potential energy surface were used to study the nucleation of bamboo-like carbon nanotubes (BCNTs). The simulations reveal that inner walls of the bamboo structure start to nucleate at the junction between the outer nanotube wall and the catalyst particle. In agreement with experimental results, the simulations show that BCNTs nucleate at higher dissolved carbon concentrations (i.e., feedstock pressures) than those where non-bamboo-like carbon nanotubes are nucleated.

[*]Corresponding author, Email: fengding@fy.chalmers.se

**Introduction**

Since the discovery of $C_{60}$[1], carbon nanostructures (including fullerenes,[1] carbon onions[2], carbon nanotubes[3] and carbon fibers[4]) have been the focus of theoretical and experimental studies due, in part, to their potential importance in materials science applications. In contrast to carbon fibers (CFs), carbon nanotubes (CNTs) are comprised of one or more concentric graphitic cylinders, which are parallel - or have a very small angle - to their symmetry axis. CNTs have been the target of many recent numerical studies because of their unique electronic, mechanical, thermal and magnetic properties.[5] The three types of CNTs, identified experimentally, are single-walled carbon nanotubes (SWNTs),[6,7] multi-walled carbon nanotubes (MWNTs)[3] and bamboo-like carbon nanotubes (BCNTs).[8]

The three experimental procedures that are most widely used to produce CNTs are the arc discharge (ARC),[6,7] laser ablation (LA)[9] and catalytic chemical vapor deposition (CCVD) methods.[10] Catalytic transition metals, such as Fe, Co, Ni and their alloys, play a crucial role in experimental production methods. Catalytic metal particles are often included in ARC and LA methods to produce SWNTs, whereas CCVD methods are used to produce all the three types CNTs. In CCVD experiments the type, quality and quantity of the products depend sensitively on the size of the catalyst particles, temperature and carbon feedstock. In certain experiments the substrate may also play an important role.

In contrast to SWNTs and MWNTs, BCNTs have regularly occurring compartment-like graphitic structures inside the nanotube (which, as the name indicates, resemble the structure of the bamboo plant). In spite of the fact that BCNTs have been studied extensively, their growth mechanism is still poorly understood. The well-known vapor-liquid-solid (VLS) model[11, 4, 12] has been extended to several other growth models, including a base growth model proposed by C.J. Lee and J. Park[13] and a tip growth model proposed by W. Z. Li *et al.*[14] These models are based on experimental observations, and there are still many unresolved issues regarding the atomic-level details of the BCNT nucleation process:

1. It has been suggested [13] that the graphitic compartment (i.e., the inner bamboo wall) nucleates from the central part of the catalyst particle inside the outer wall (see region A in Fig. 1). This is at the opposite side of the catalyst particle from where carbon atoms are deposited (i.e., from where the feedstock decomposes to release carbon atoms into the catalyst particle – region B in Fig. 1). It is clear that bulk diffusion of carbon atoms is required for them to reach region A from B, since surface diffusion is prevented by the presence of the outer wall. It is not clear why carbon atoms precipitate and nucleate at the central part inside the outer wall (region A) instead of, for example, in regions closer to the junction between the outer wall and the metal particle surface.

2. Experimental results show that BCNTs nucleate at higher carbon feedstock pressures than those required for SWNT and MWNT nucleation.[14] The (atomic-level) reasons for this are not yet fully understood.

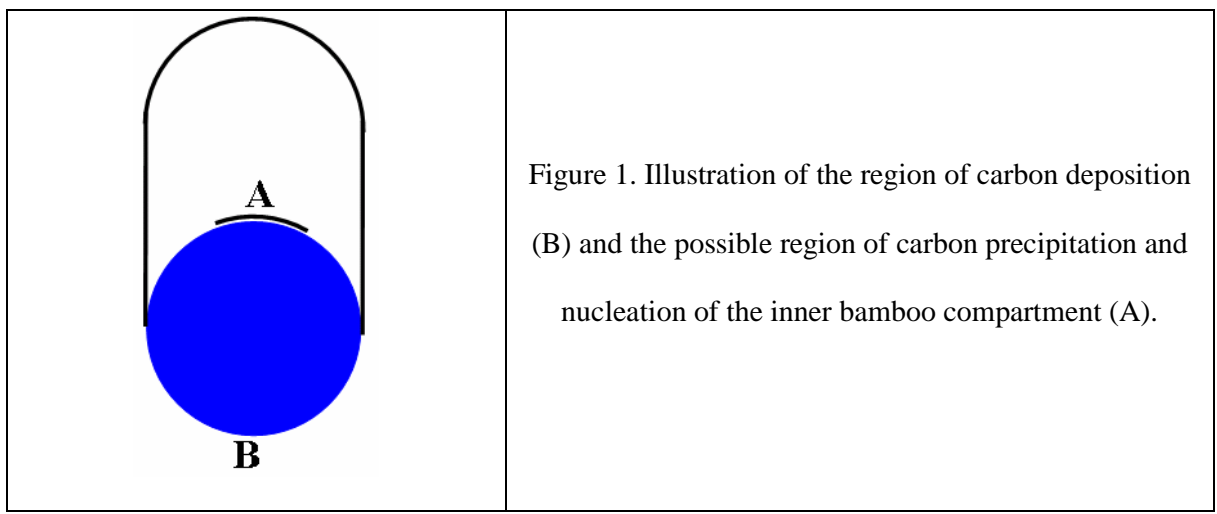

Figure 1. Illustration of the region of carbon deposition (B) and the possible region of carbon precipitation and nucleation of the inner bamboo compartment (A).

This contribution presents the results of a molecular dynamics study of the BCNT nucleation mechanism. It addresses, and provides possible solutions, to the two issues listed above.

**Simulation methods**

Theoretical studies, which may be static calculations[15] or dynamics simulations[16, 17], provide a valuable complement to experimental investigations since they provide a detailed, atomic-level understanding of CNT nucleation mechanisms. Static calculations are done at zero Kelvin (i.e., minimum energy geometries are often determined and analyzed), and hence they do not directly include entropic effects that may be important under CNT nucleation temperatures (usually higher than 823 K). Molecular dynamics (MD) simulations are usually performed at elevated temperatures and, in this respect, may provide information that is more relevant for the experimental growth conditions. However, since interatomic forces for several thousands or even millions of geometries are required for MD, it is computationally more expensive than static calculations. For example, a density functional theory (DFT) based MD study (i.e., a direct dynamics study), which explicitly includes the change in electron density over time, allowed for the simulation of only several tens of picoseconds and a limited number of atoms (about 100-200).[17] This is not sufficient to simulate the entire nucleation process. In contrast, MD simulations based empirical potential energy surfaces (PESs) may have lower quantitative accuracy than DFT based MD, but they allow for long simulations (up to 100 ns) of much larger systems (up to thousands of atoms).[16] In this way empirical PES based MD simulations allow for a systematic study of the trends and qualitative mechanisms of CNT nucleation under conditions that are relevant to experimental growth techniques.

We have previously developed an empirical PES to study iron-catalyzed nucleation of CNTs for a range of catalyst particle sizes, temperatures and carbon deposition rates.[18, 19, 20] This PES was developed from Fe-Fe, Fe-C and C-C potential functions and interactions strengths that are available in the literature and that were obtained from experimental and ab initio data.[18] The simulations, which show that SWNT nucleation on iron catalyst particles occurs between 800 K and 1400 K (which is the same as the interval used in CCVD experiments) revealed the detailed SWNT growth mechanism and identified critical stages in the nucleation process.[18, 19, 20] The PES also reproduces the correct trends in the iron-carbide phase diagram (which may be important if the VLS model is correct) and the correct decrease in melting point with decreasing cluster size.[21, 22]

## Results and Discussion

Fig. 2 shows snapshots during the nucleation of a bamboo-like CNT on an $Fe_{100}$ particle. The iron particle initially consists only of Fe atoms (Fig. 2a), and is thermalized to the desired simulation temperature (1000 K for the simulation shown in Fig. 2). Carbon atoms are subsequently deposited on the metal particle at a rate of one every 40 ps (by depositing C atoms on the particle we circumvent the need to model the catalytic decomposition of feedstock gas that cannot be studied using our empirical PES). Initially all of the deposited carbon atoms dissolve into the catalyst particle, but once the particle is supersaturated in carbon, the carbon atoms start to precipitate on the particle surface (panel b). However, these precipitated atoms are not stable on the particle surface and may dissolve back into the catalyst particle. Further increase in the number of dissolved atoms over time (which is associated with a highly supersaturated particle) results in a larger number of precipitated atoms, which can form C-C bonds with neighboring atoms. This results in the formation of carbon strings and polygons that are stable on the particle surface (panel c). These strings and polygons provide nucleation sites for further carbon precipitation, and hence grow into carbon islands (panel d). The interaction between the C atoms at the centre of the island (which are sp2 carbons) and the metal particle are fairly weak, and at elevated temperatures (T>800 K) the kinetic energy is sufficient to overcome the attractive forces between these atoms and the particle and the carbon island lifts off the particle surface to form a graphitic cap (panel e). The stronger interactions between the C atoms at the edge of the island (which have dangling bonds) and the particle anchors the cap onto the particle, and carbon atoms that continue to precipitate on the surface are incorporated at the edges of the cap. In this way the cap grows bigger and eventually forms a SWNT or the outer wall of a BCNT (panel f and g).

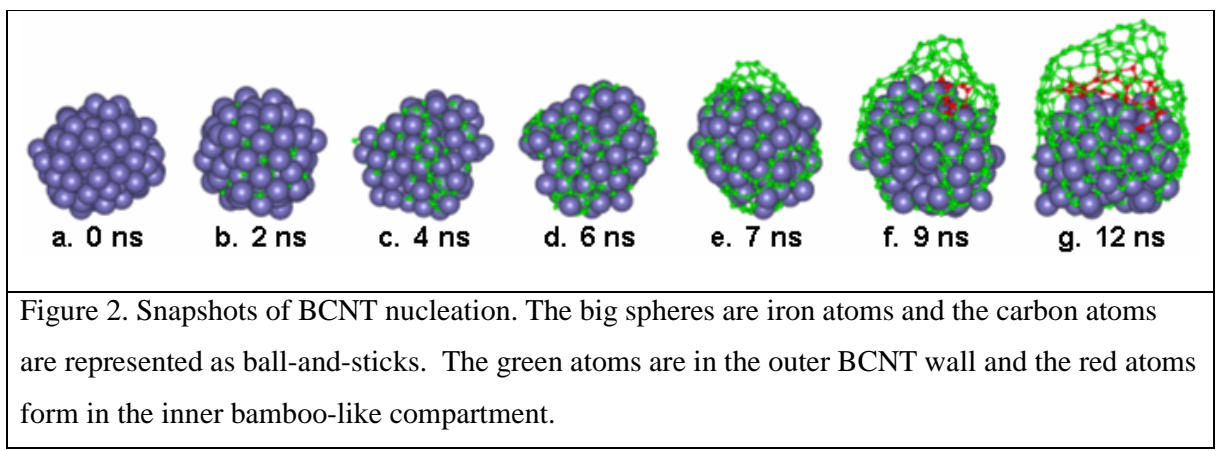

Figure 2. Snapshots of BCNT nucleation. The big spheres are iron atoms and the carbon atoms are represented as ball-and-sticks. The green atoms are in the outer BCNT wall and the red atoms form in the inner bamboo-like compartment.

The initial stages in the formation of the bamboo compartment inside the outer wall are shown in Figs. 2f and g (red atoms). It is clear from the structure in panel f that the inner compartment nucleates from regions near the outer wall (i.e., on one side of the metal particle) and not from a central region (e.g., A in Fig. 1). That is, it is evident from the simulations that carbon atoms, which precipitate on the particle surface in the inside the outer tube, are only stable if they interact with the carbon atoms in the outer wall. Hence, it is only these carbon atoms, located near the 'corners' between the particle surface and the outer wall, that are sufficiently stable to nucleate the inner compartment (panel g).

Further understanding of the CNT growth mechanism can be obtained by comparing possible sites where carbon atoms can precipitate on the cluster surface. Four possible sites, A-D, are shown in Fig 3a. Sites A and B are outside the first (or outer) CNT wall, whereas sites C and D are on the particle surface that is on the inside of the first CNT wall. Also, sites A and D are close to the junction between the CNT wall and the particle surface whereas B and C are far from this junction. As discussed above, carbon atoms that precipitate at sites B and C are not as stable as atoms precipitated at sites A and D, since the atoms at the latter sites are stabilized by their interactions with the CNT wall. This is expected since the binding energy of a single carbon atom inside a SWNT is about 1-2 eV[23], and hence the outer wall stabilizes atoms that precipitate in its vicinity (e.g., at D). Also, carbon atoms that are deposited on the surface (in region B) can arrive at site A by surface or bulk diffusion, but bulk diffusion is required for them to precipitate at D (or C). Hence, precipitation at site A is expected to be more rapid (and more stable since these atoms join to the dangling carbon bonds at the CNT end) and hence incorporation of carbon atoms into a SWNT structure (site A) is favored over the formation of a second wall or compartment structure at D. This is consistent with the experimental observations that non-bamboo-like CNTs growth dominates over BCNT at lower feedstock pressures (i.e., lower carbon deposition rates).[14] This is also seen in our simulations.

At larger carbon deposition rates there is a larger concentration of dissolved carbon in the metal particle. Under these conditions the rate of precipitation of atoms (at all surface sites) increases, and there is an increased probability that atoms that precipitate at site D can nucleate the inner bamboo compartment. That is, an increase in the rate of carbon atom precipitation at site D (when the carbon deposition rate is high) increases the probability that a second, compartment-like structure inside the outer wall is formed. This is also in agreement with experimental observations that BCNTs grow at higher feedstock pressures than those used for SWNT and MWNT growth.[14] Once the bamboo compartment is nucleated at site D, additional carbon atoms can join to this structure to form the complete bamboo structure as illustrated in Figs 3b and c.

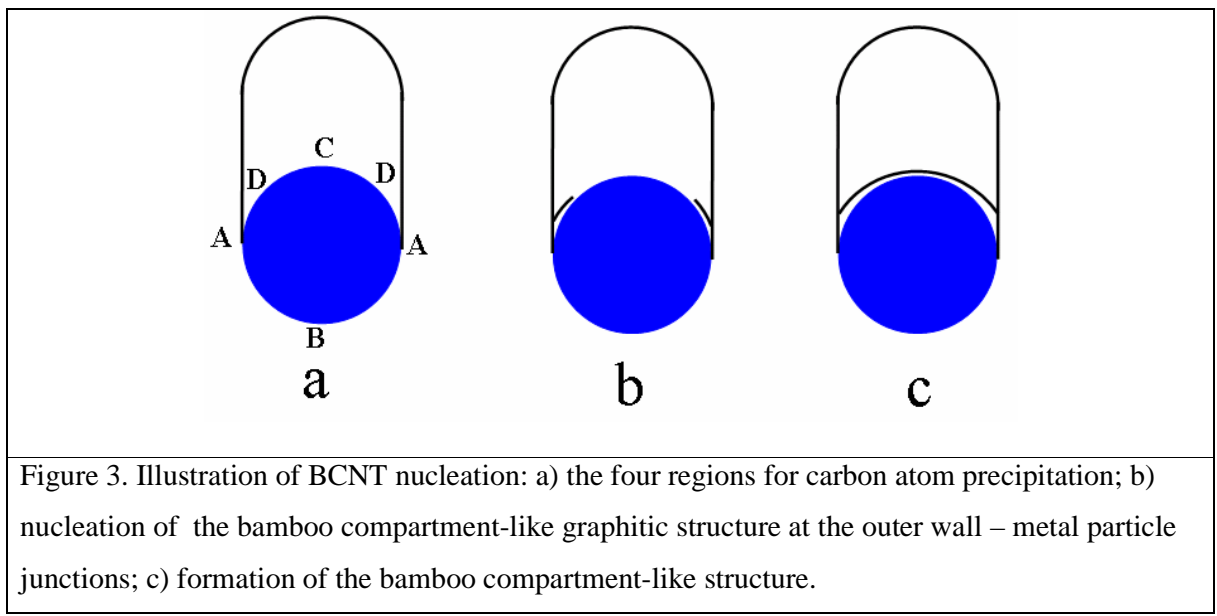

Figure 3. Illustration of BCNT nucleation: a) the four regions for carbon atom precipitation; b) nucleation of the bamboo compartment-like graphitic structure at the outer wall – metal particle junctions; c) formation of the bamboo compartment-like structure.

## Conclusion

Molecular dynamics studies have been used to study the detailed mechanism of bamboo-like carbon nanotube (BCNT) nucleation on an $Fe_{100}$ catalyst particle. The nucleation of the bamboo compartment-like structure inside the outer CNT starts at the junction between the outer wall and the metal particle surface. In this region the precipitated carbon atom is stabilized by the outer wall (compared to atoms that precipitate far away from the outer wall). In addition, since precipitation of carbon atoms inside the outer wall can only be achieved by bulk diffusion, formation of BCNTs only occurs at high dissolved carbon concentrations (i.e., high feedstock pressures). At lower dissolved carbon concentrations surface diffusion to, and precipitation at, regions outside the outer wall dominates, and this leads to SWNT growth. This is in agreement with experimental results which show that BCNTs are formed at higher carbon concentrations than those required for SWNT and MWNT formation.

## References:

[1] H.W. Kroto, J.R. Heath, S.C. O'Brien, R.F. Curl, R.E. Smalley, Nature 318 , 162, (1985)

[2] D. Ugarte, Nature 359,707 (1992)

[3] Iijima, S. Nature, 354, 56 (1991)

[4] Baker RTK, Baker MA, Harris PS, Feates FS, Waite RJ. J Catal, 26, 51 (1972)

[5] H. Dai, Surface Science 500, 218 (2002).

[6] S. Iijima, T. Ichihashi, Nature 363, 603 (1993).

[7] D. S. Bethune et al., Nature 365, 603 (1993).

[8] Y. Saito and T. Yoshikawa, J. Cryst. Growth, 134, 154 (1993)

[9] A. Thess et al., J. Fischer, R. Smalley, Science 273, 483 (1996)

[10] J. Kong, H. Soh, A. Cassell, H. Dai, Nature 395, 878 (1998).

[11] Wagner RS, Ellis WC.. Appl Phys Lett, 4, 89. (1964)

[12] Saito Y. Carbon, 33 979 (1995)

[13] C. J. Lee and J. Park, App. Phys. Lett., 77, 3397 (2000)

[14] W. Z. Li, J. G. Wen, Y. Tu and Z. F. Ren, Appl. Phys. A., 73, 259 (2001)

[15] X.Fan, R.Buczko, A.A.Puretzky, D.B.Geohegan, J.Y.Howe, S.T.Pantelides, S.J.Pennycook, Phys.Rev.Lett. 90, 145501 (2003)

[16] Y.Shibuta, S.Maruyama, Physica B 323, 187 (2002)

[17] J.Gavillet, A.Loiseau, C.Journet, F.Willaime, F.Ducastelle, J.C.Charlier, Phys.Rev.Lett. 87, 275504 (2001)

[18] F. Ding, A. Rosén, and K. Bolton, J. Phys. Chem. B, 108, 17369 (2004).

[19] F. Ding, A. Rosen, and K. Bolton, Chem. Phys. Lett. 393, 309 (2004)

[20] F. Ding, A. Rosén, and K. Bolton, J. Chem. Phys. 121, 2775 (2004).

[21] F. Ding, K. Bolton and A. Rosén. J. Vac. Sci. Technol. A **22**, 1471 (2004).

[22] F. Ding, A. Rosén, and K Bolton, Phys. Rev. B 70, 075416 (2004).

[23] A. V. Krasheninnikov, K. Nordlund, P. O. Lehtinen, A. S. Foster, A. Ayuela, R. M. Nieminen, Carbon 42 1021 (2004)